\documentclass[12pt]{JHEP3}

\preprint{  }

\usepackage{epsfig,multicol}
\usepackage{amsmath}
\usepackage{graphics}
\usepackage{graphicx}
\usepackage{dcolumn}
\usepackage{amssymb}
\usepackage{amsthm}
\usepackage{amsfonts}
\usepackage{subfigure}
\usepackage{setspace}

\title{Holographic entanglement entropy in insulator/superconductor transition with Born-Infeld electrodynamics}
\author{Weiping Yao\footnote{Email: yao11a@126.com}  and
Jiliang  Jing\footnote{Corresponding author. 
jljing@hunnu.edu.cn}
\\ Department of Physics, and Key Laboratory of Low Dimensional Quantum Structures and Quantum Control of Ministry of Education, Hunan Normal University, Changsha, Hunan 410081, P. R. China}

\abstract{We investigate the holographic entanglement entropy in the
insulator/superconductor phase transition for the Born-Infeld
electrodynamics with full backreaction in five-dimensional AdS
soliton spacetime. We note that the holographic entanglement entropy
is a good probe to study the properties of the phase transition, and
the Born-Infeld factor $b$ has no effect on the critical chemical
potential $\mu_c$. We find that both in the half space and the belt
one, the non-monotonic behavior of the entanglement entropy versus
the chemical potential is a general property, and the entanglement
entropy increases with the increase of the Born-Infeld factor in the
superconductor phase. Particularly, there exists
confinement/deconfinement phase transition in the strip geometry and
the critical width $\ell_c$ is dependent of the Born-Infeld
parameter. }

\keywords{holographic entanglement entropy, AdS/CFT correspondence, Born-Infeld electrodynamics}

\begin{document}

\section{Introduction}

The anti-de Sitter/conformal field theories (AdS/CFT) correspondence
\cite{J.Maldacena1998, EWritten1998, S.S.Gubser1998} has provided us
a useful theoretical method to study the strongly coupled systems in
various fields of physics. In recent years, it has been widely
applied to study the holographic model of superconductors
\cite{S.A.Hatnall2009, C.P.Herzog2009, G.T.horowitz2010}. It is
indicated that the bulk AdS black hole becomes unstable and scalar
hair condenses as one tunes the temperature of black hole. Besides
the bulk AdS black hole spacetime, since the AdS soliton
configuration has the same boundary topology as the Ricci flat black
hole and the AdS space in the Poincar\'{e} coordinates, a
superconductor phase dual to an AdS soliton configuration without
backreaction was studied in Ref. \cite{T. Nishioka2010}. It was
shown that if one increases a chemical potential $\mu$ to the AdS
soliton, there is a second order phase transition at a critical
value $\mu_c$ beyond which the charge scalar field turns on, even at
zero temperature. Considering the backreaction of the matter fields
on the soliton geometry, a first order phase transition occurs for
the same chemical potential as the backreaction is strong enough
\cite{G.T. Horowitz2010}. When thinking about the higher order
correction to the Maxwell field, it is of interest to investigate
the effect of the Born-Infeld electrodynamics \cite{M.Born1934,
G.W.Gibbons1995, B.Hoffmann1935, W.Heisenberg1936,  Oliveira1994,
O.Miskovic2011} on the insulator/superconductor transition. The
Born-Infeld electrodynamics, which was proposed in 1934 to avoid the
infinite self-energies for charged point particles arising in
Maxwell theory \cite{M.Born1934}, displays good physical properties
including the absence of shock waves and birefringence. It was found
that the Born-Infeld electrodynamics is the only possible non-linear
version of electrodynamics that is invariant under electromagnetic
duality transformation \cite{G.W.Gibbons1995}. The Lagrangian
density for Born-Infeld theory is
$L_{BI}=\frac{1}{b^2}\left(1-\sqrt{1+\frac{b^2F^{ab}F_{ab}}{2}}\right)$
with $F^2=F_{\mu\nu}F^{\mu\nu}$ and this Lagrangian will reduce to
the Maxwell case as the coupling parameter $b$ approaches zero. In
the limit of the probe approximation, the Born-Infeld factor $b$ has
nothing to do with the insulator/superconductor transition
\cite{xixu zhao2012}. In this paper we will extend the previous
works by considering the full backreaction of the matter fields with
the Born-Infeld electrodynamics on the soliton geometry.

On the other hand, the entanglement entropy has become a useful
quantity for studying some properties in quantum field theories and
in many-body physics recently \cite{P. Calabrese2004, Ryu2006, M.
Levin2006, A. Kitaev2006, P. Calabrese2006, L. Amico2008, B.
Hsu2009}. As is well known, dividing a given quantum system into a
subsystem $\mathcal{A}$ and its complement, the entanglement entropy
of $\mathcal{A}$ is defined as the von Neumann entropy. In quantum
many-body physics, the entanglement entropy allows one to
distinguish new topological phases and characterize critical points.
In the light of the AdS/CFT correspondence,  the entanglement
entropy may provide us new insights into the quantum structure of
spacetime \cite{M. Van Raamsdonk2009, M. Van Raamsdonk2010}. Ryu and
Takayanagi \cite{Ryu:2006bv, Ryu:2006ef} have  provided a proposal
to compute the entanglement entropy of CFTs from the minimal area
surface in gravity side.  This proposal provides a simple and
elegant way to calculate the entanglement entropy of a strongly
coupled system which has a gravity dual. Then, a lot of works have
been carried out for investigating the entanglement entropy in
various gravity theories
\cite{Nishioka:2009un,Albash:2011nq,Myers:2012ed,deBoer:2011wk,Hung:2011xb,Nishioka:2006gr,
Klebanov:2007ws,Pakman:2008ui,Ogawa:2011fw}. Refs.
\cite{Nishioka:2006gr,Klebanov:2007ws,Pakman:2008ui} presented the
calculations of the entanglement entropy for the AdS soliton
geometry. Ref. \cite{Ogawa:2011fw} considered the case with higher
derivative corrections and studied the holographic entanglement
entropy in Gauss-Bonnet gravity. Ref. \cite{Xi Dong2013} studied the
holographic entanglement entropy for general higher derivative
gravity and proposed a general formula for calculating the
entanglement entropy in theories dual to higher derivative gravity.
Since the entanglement entropy behaves like the thermal entropy of
background black holes, it can indicate not only the appearance, but
also the order of the phase transition \cite{T. Albash2012, R. G.
Cai2012, Li-Fang Li2013}. The authors \cite{Cai1203} investigated
the behavior of the entanglement entropy  in a simple holographic
insulator/superconductor model at zero temperature, and found that
the entanglement entropy as a function of chemical potential is not
monotonic in the superconductor phase. Precisely, the entanglement
entropy first increases and reaches its maximum at a certain
chemical potential and then decreases monotonically as chemical
potential increases. This non-monotonic behavior is quite different
from the case of the metal/superconductor phase transition \cite{T.
Albash2012}. Furthermore, this non-monotonic behavior of the
entanglement entropy versus the chemical potential still stands in
the St\"{u}ckelberg holographic insulator/superconductor model
\cite{Cai1209}. Motivated by the study of the entanglement entropy
in the higher correction to gravity and in the
insulator/superconductor transition, it is natural to study how the
entanglement entropy will be modified as a result of the correction
to the Maxwell field. Here we would like to study the entanglement
entropy in the insulator/superconductor phase transition with
Born-Infeld electrodynamics in the two geometry configurations which
are descried by the half space and the strip one, respectively. We
find that the non-monotonic behavior of the entanglement entropy
versus the chemical potential is universal in this model. And the
entanglement entropy increases as the Born-Infeld factor $b$
increases in the superconductor phase. Particularly, the
confinement/deconfinement phase transition exists in the strip
geometry.

The framework of this paper is as follows. In Sec. II, we will
introduce the holographic superconductor models and derive the
equations of motions.  In Sec. III, we will study the phase
transition in the AdS soliton gravity with Born-Infeld
electrodynamics. In Sec. IV, we will calculate the holographic
entanglement entropy for the half geometry and strip space
respectively. In Sec. V,  we will conclude our main results of this
paper.

\section{Equations of motion and boundary conditions}
Let us begin with the action for a Born-Infeld electromagnetic field coupling
with a charged scalar field with a
negative cosmological constant in five-dimensional spacetime
\begin{eqnarray}\label{action}
S&=&\int d^5
x\sqrt{-g}\left(R+\frac{12}{L^2}\right)
+\int d^5 x\sqrt{-g}\left[\frac{1}{b^2}\left(1-\sqrt{1+\frac{b^2
F^{ab}F_{ab}}{2}}\right)\right.\nonumber \\
&&\left.-|\nabla\Psi-i q A\Psi
|^2-m^2|\Psi|^2\right],
\end{eqnarray}
where $L$ is the radius of AdS spacetime, $g$ is the determinant of
the metric,  $q$ and $m$ are respectively the charge and the mass of
the scalar field, and $b$ is the Born-Infeld coupling parameter.
Since we are interested in including the backreaction, we will
choose the ansatz of the geometry of the AdS soliton with the form
\begin{equation}
ds^2
=\frac{dr^2}{r^2B(r)}+r^2\left[-e^{C(r)}dt^2+dx^2+dy^2+e^{A(r)}B(r)d\chi^2\right].
\end{equation}
Without loss of generality, we set $L=1$ in this paper. As in the
usual AdS soliton, to get a smooth geometry at the tip $r_0$, $\chi$
should be periodic with the period
\begin{equation}
\Gamma=\frac{4\pi e^{-A(r_0)/2}}{r_0^2B'(r_0)},
\end{equation}
and $B(r)$ vanishes at the tip of the soliton. The electromagnetic
field and the scalar field can be taken as
 \begin{equation}
 A_t=\phi(r),\ \ \psi=\psi(r).
 \end{equation}
The equations of motion under the above ansatz can be obtained as follows
\begin{eqnarray}
&&
\psi''+\left(\frac{5}{r}+\frac{A'}{2}+\frac{B'}{B}+\frac{C'}{2}\right)\psi'+\frac{1}{r^2B}
\left(\frac{e^{-C}q^2\phi^2}{r^2}-m^2\right)\psi=0\;, \\
&& \phi''+\left(\frac{3}{r}+\frac{A'}{2}+\frac{B'}{B}
-\frac{C'}{2}\right)\phi'-b^2e^{-C}B\left(\frac{3}{r}+\frac{A'}{2}+\frac{B'}{2B}
\right)\phi'^3\nonumber \\ &&\ \ \ \ -
\frac{2\psi^2q^2\phi}{r^2B}\left(1-b^2
e^{-C}B\phi'^2\right)^{\frac{3}{2}}=0\;,
\\
&& A'=\frac{2r^2C''+r^2C'^2+4rC'+4r^2\psi'^2-2e^{-C}\phi'^2(1-b^2
e^{-C}B\phi'^2)^{-\frac{1}{2}}}{r(6+rC')}\;, \label{Aeq}
 \\
 &&
C''+\frac{1}{2}C'^2+\left(\frac{5}{r}+\frac{A'}{2}+\frac{B'}{B}\right)C'
-\left[\phi'^2(1-b^2 e^{-C}B\phi'^2)^{-\frac{1}{2}}\right.\nonumber \\
&& \left. \ \ \ \ +\frac{2q^2
\phi^2\psi^2}{r^2B}\right]\frac{e^{-C}}{r^2} =0,
\\
&&
B'\left(\frac{3}{r}-\frac{C'}{2}\right)+B\left[\psi'^2-\frac{1}{2}
A'C'+\frac{1-(1-b^2 e^{-C}B\phi'^2)^{\frac{1}{2}}}{r^2b^2B(1-b^2
e^{-C}B\phi'^2)^{\frac{1}{2}}}+\frac{12}{r^2}
\right] \nonumber \\
&&
\ \ \ \ +\frac{1}{r^2}\left(\frac{e^{-C}q^2\phi^2\psi^2}{r^2}+m^2\psi^2-12\right)
=0,
\end{eqnarray}
where a prime denotes the derivative with respect to $r$. By
considering a series solution about the tip of the soliton
($r=r_{0}$) and using the boundary condition $B(r_{0})=0$, we get
four independent parameters, i.e., $r_0,$ $\psi(r_0),$ $\phi(r_0)$
and $ C(r_0)$. Interestingly, we note that the above equations of
motion have useful scaling symmetries
\begin{eqnarray}
&& r\rightarrow \alpha r,\qquad (\chi,x,y,t)\rightarrow(\chi,x,y,t)/\alpha,\qquad\phi\rightarrow \alpha\phi\label{scaling1},\\
&& C\rightarrow C-2\ln{\beta},\qquad t\rightarrow \beta t,\qquad\phi\rightarrow \phi/\beta.\label{scaling2}
\end{eqnarray}
Therefore, we can pick any values for the position of the tip $r_0$ and $C(r_0)$. Here, we take $r_0=1,$ $C(r_0)=0$ for simplicity.

At the spatial infinity $(r\rightarrow\infty)$, as we want the
spacetime to be asymptotically AdS, the matter fields have the form
\begin{eqnarray}
\phi\sim\mu-\frac{\rho}{r^{2}}\label{b:infinityphi},\\
\psi\sim\frac{\psi_{-}}{r^{\Delta_{-}}}+\frac{\psi_{+}}{r^{\Delta_{+}}},
\label{b:infinitypsi}
\end{eqnarray}
where the conformal dimensions of the operators are $\Delta_{\pm}=2
\pm\sqrt{4+m^2}$, $\mu$ and $\rho$ are the corresponding chemical
potential and charge density in the dual field theory, respectively.
According to the AdS/CFT correspondence, the coefficients $\psi_-$
and $\psi_+$ correspond to the vacuum expectation values
$\psi_{-}=<\mathcal{O}_{-}>$,  $\psi_{+}=<\mathcal{O}_{+}>$ of an
operator $\mathcal{O}$ dual to the scalar field. For the sake of
obtaining the stability in the asymptotic AdS region, we can impose
boundary conditions that either $\psi_-$ or $\psi_+$ vanishes. In
addition, in five-dimensional spacetime, when $-4 < m^2<-3$, the
scalar field admits two different quantization related by a Legendre
transform \cite{I. R. Klebanov1999}. Hence, we will focus on the
case $m^2=-15/4$, $q=2$ and  $\psi_-=0$ in the following
calculation. Furthermore, to recover the pure AdS boundary, we also
need $A(r\rightarrow \infty)=0$ and $C(r\rightarrow \infty)=0$.

Using the scaling symmetries (\ref{scaling1}), we can rescale the
quantities as
\begin{equation}\label{scaling3}
\Gamma\rightarrow\frac{1}{\alpha}\Gamma,\ \ \mu\rightarrow\alpha\mu,
\ \ \rho\rightarrow\alpha^3\rho,\ \
\langle\hat{O}_2\rangle\rightarrow\alpha^{\frac{5}{2}}\langle\hat{O}_2\rangle.
\end{equation}
In the next section, we will be centered on the following
dimensionless quantities $~~\mu\Gamma, ~~\rho\Gamma^3~~$ and
$~~\langle\hat{O}_2\rangle^{\frac{2}{5}}\Gamma$.

\section{Insulator/Superconductor Phase Transition}

To study the phase transition for Born-Infeld electrodynamics with full backreaction in the five-dimensional AdS soliton spacetime,
we make a coordinate transformation from $r-$coordinate to $z-$coordinate
by defining $z=r_0/r$. Then, the equations of motion can be rewritten as
\begin{eqnarray}
&&
\psi''+\left(\frac{A'}{2}+\frac{B'}{B}+\frac{C'}{2}-\frac{3}{z}
\right)\psi'+\frac{1}{z^2B}
\left(e^{-C}q^2\phi^2z^2-m^2\right)\psi=0\;, \\
&& \phi''+\left(\frac{A'}{2}+\frac{B'}{B}
-\frac{C'}{2}-\frac{1}{z}\right)\phi'-b^2z^4e^{-C}B
\left(\frac{A'}{2}+\frac{B'}{2B} -\frac{3}{z}\right)\phi'^3\nonumber
\\ &&\ \ \ \ - \frac{2\psi^2q^2\phi}{z^2B}\left(1-b^2z^4
e^{-C}B\phi'^2\right)^{\frac{3}{2}}=0\;,
\\
&& A'=-\frac{z[2C''+C'^2+4\psi'^2-2z^2e^{-C}\phi'^2(1-b^2z^4
e^{-C}B\phi'^2)^{-\frac{1}{2}}]}{(6-zC')}\;, \label{Aeq1}
\\
 &&
C''+\frac{C'^2}{2}+\left(\frac{A'}{2}+\frac{B'}{B}-\frac{3}{z}
\right)C'-\left[z^2\phi'^2(1-b^2 z^4
e^{-C}B\phi'^2)^{-\frac{1}{2}}\right. \nonumber \\ && \left.\ \ \ \
+\frac{2q^2 \phi^2\psi^2}{B}\right]e^{-C} =0,
\\
&&
B'\left(\frac{3}{z}+\frac{C'}{2}\right)-B\left[\psi'^2-\frac{1}{2}
A'C'+\frac{1-(1-b^2z^4
e^{-C}B\phi'^2)^{\frac{1}{2}}}{z^2b^2B(1-b^2z^4
e^{-C}B\phi'^2)^{\frac{1}{2}}}+\frac{12}{z^2}
\right] \nonumber \\
&& \ \ \ \
-\frac{1}{z^2}\left(e^{-C}q^2z^2\phi^2\psi^2+m^2\psi^2-12\right)
=0,
\end{eqnarray}
where the prime now denotes a derivative with respect to $z$. The
region $r_0<r<\infty$ now corresponds to $1>z>0$. From above
discussion, by choosing $\phi(r_0)$ as a shooting parameter, we can
solve the equations of motion for the given $m^2,~q,~\psi(r_0)$.
After solving the equations of motion, according to the AdS/CFT
correspondence, we can get the chemical potential $\mu$ and the
charge density $\rho$ from the asymptotic behavior of $\phi$ through
Eq. (\ref{b:infinityphi}). We can also get the vacuum expectation
value of the scalar operator ($<\mathcal{O}>=\psi_{+}$) from Eq.
(\ref{b:infinitypsi}). For each choice of $\psi(r_0)$, we can solve
the equations of motion for $\psi,\phi,A,B,C$. Here, we present the
solutions of these equations for the factor~$\psi_0=1.5$ by
figures. In Fig.~\ref{function}, $\psi(z)$ and $\phi(z)$ are the
scalar field and static electric potential, respectively, and $A(z)$
and $B(z)$ are two metric functions. For the fixed $z$, we can see
that the functions have different values with the increase of the
Born-Infeld factor $b$. So it is of interest to study the effect of

the Born-Infeld factor $b$ on the phase transition in this system.
\FIGURE{
\includegraphics[scale=0.3]{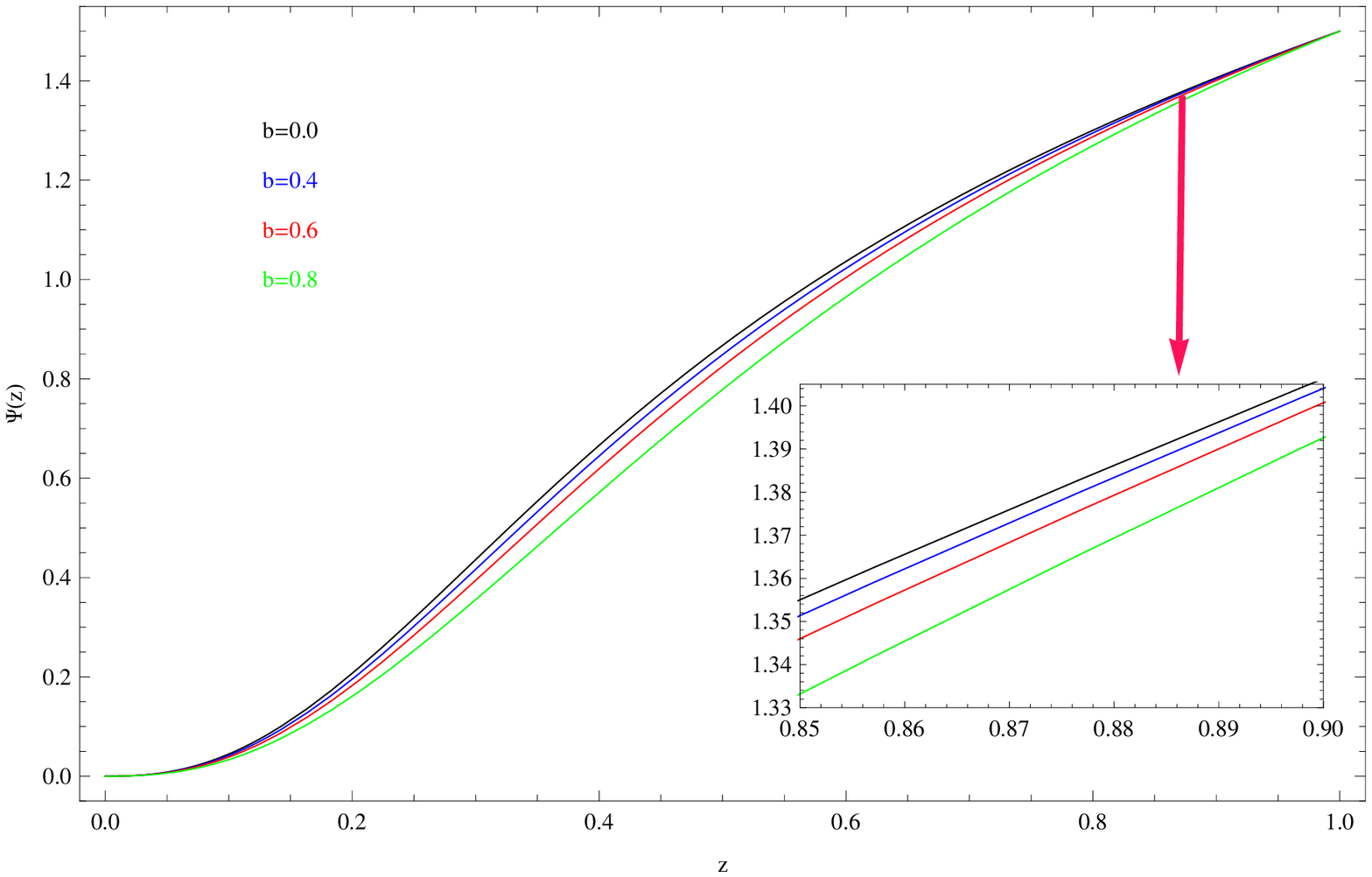}
\includegraphics[scale=0.31]{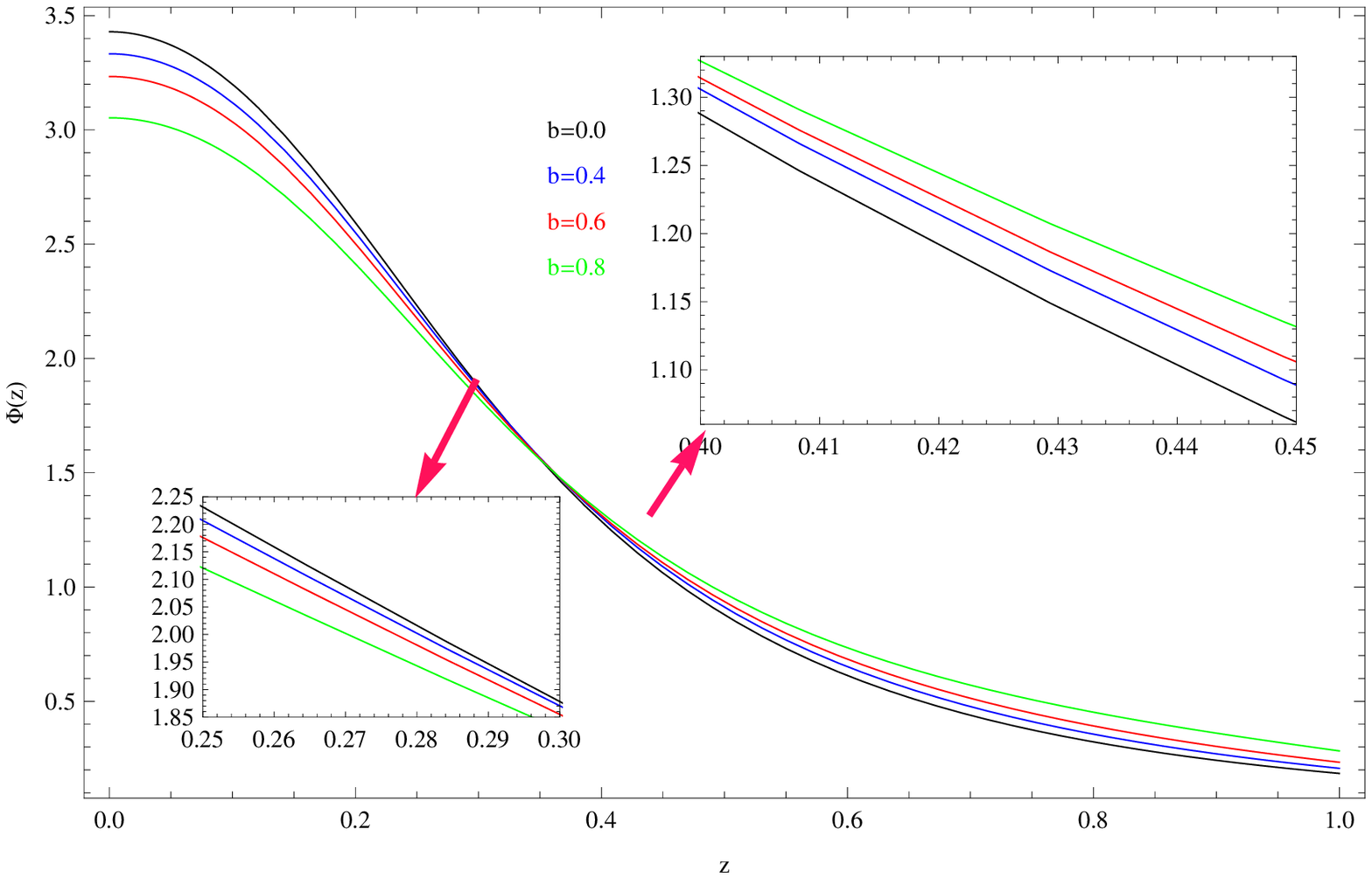}
\includegraphics[scale=0.3]{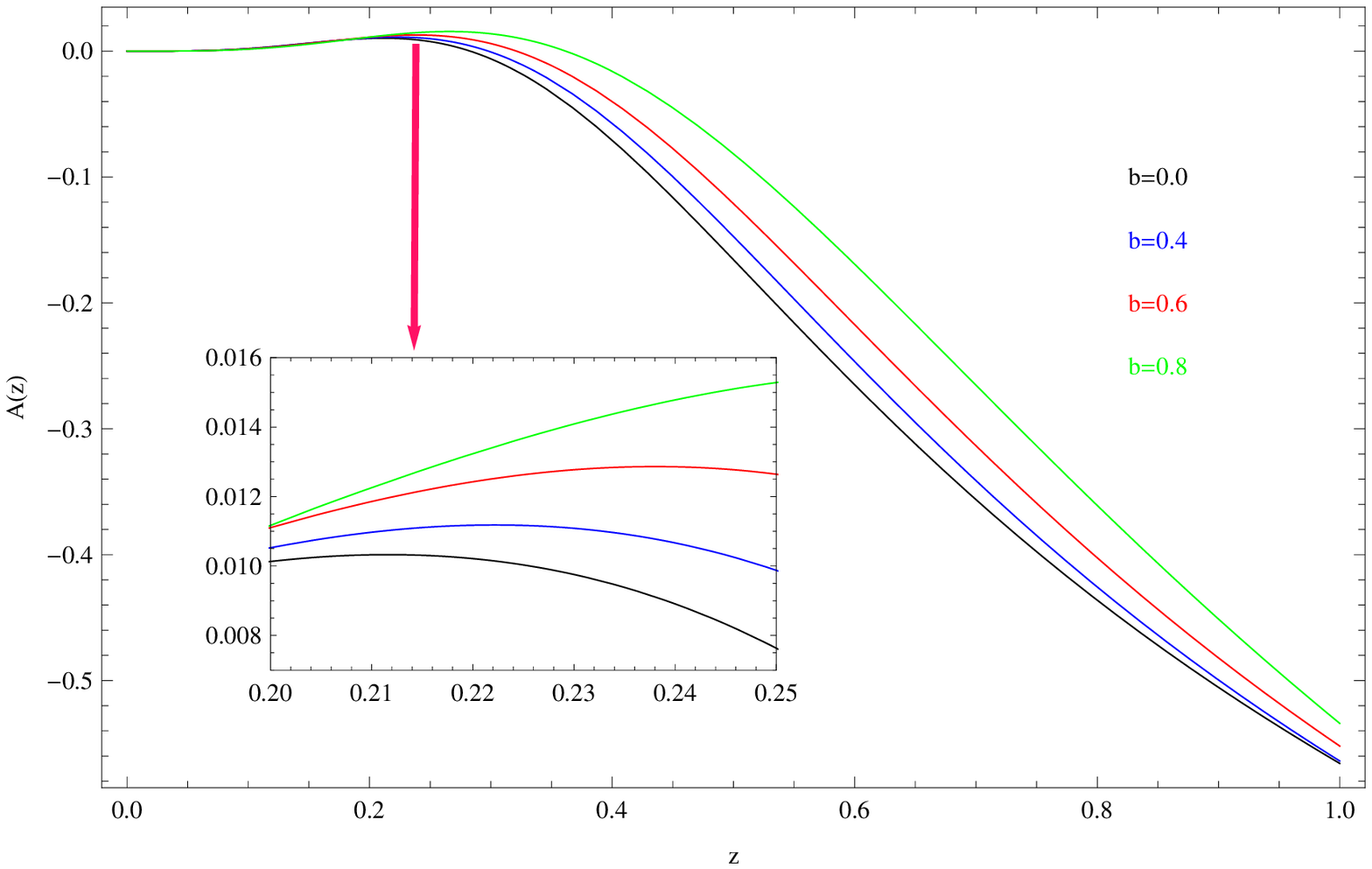}
\includegraphics[scale=0.32]{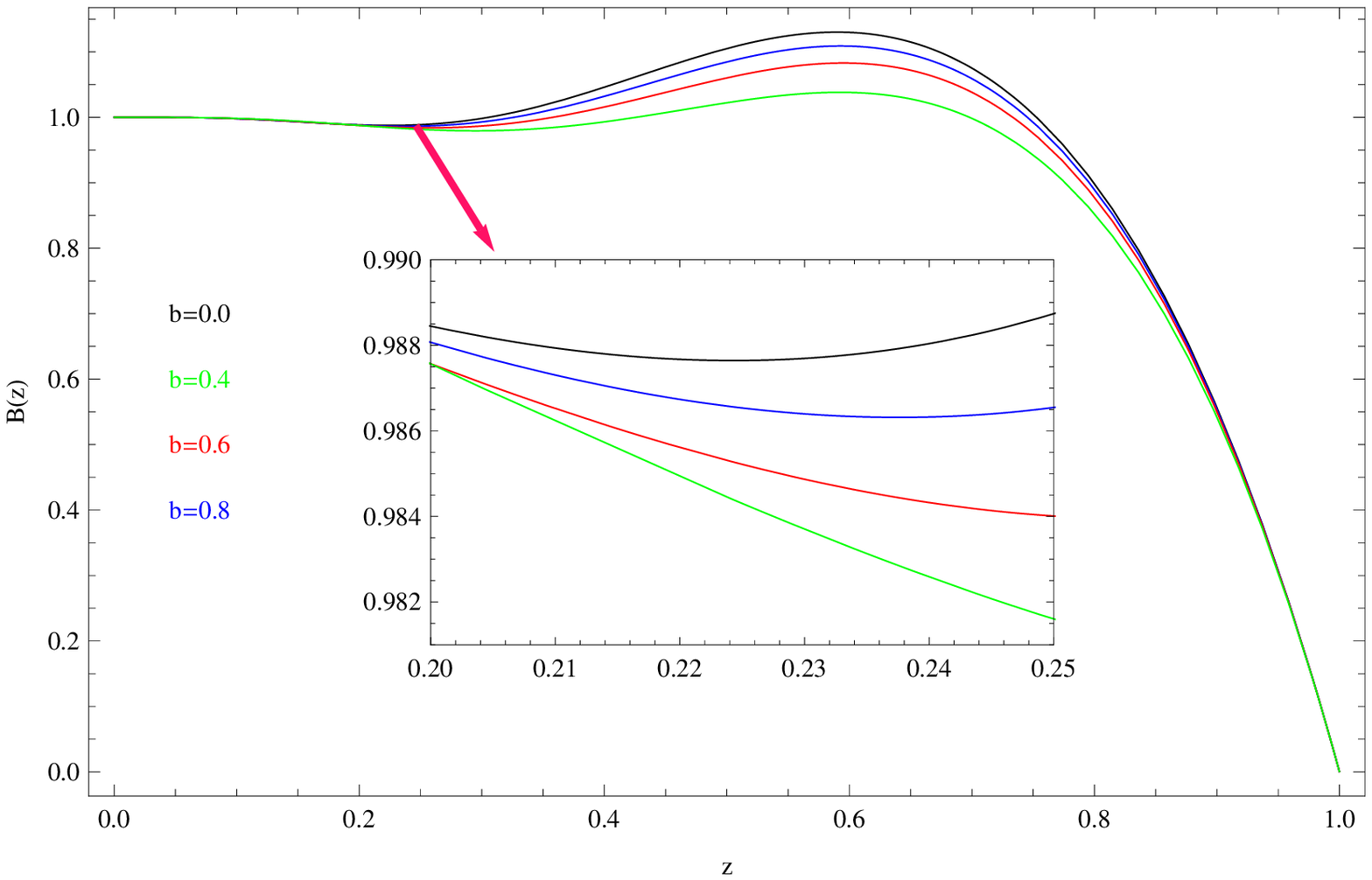}
\caption{\label{function} A typical soliton solution with
nonvanishing scalar hair for the different values of parameter $b$.
Here the value of the scalar field at the tip is $\psi_0=1.5$. The
four lines from bottom to top correspond to increasing Born-Infeld
factor, i.e.,  b = 0 (black), 0.4 (blue), 0.6 (red) and 0.8 (green),
and the corresponding identifications
are~$\Gamma\simeq2.4478, \Gamma\simeq 2.4455, \Gamma\simeq2.4310~ and~ \Gamma\simeq2.4092$,
respectively.} }
However, the solutions obtained in this way have
different periods $\Gamma$ for the $\chi$-coordinate. With the aim
of comparing different solutions for the same boundary behavior, we
can use the symmetry (\ref{scaling3}) to set all of the periods
$\Gamma$ equal. As the identification length $\Gamma$ in the pure
soliton is $\pi$, we will scale all $\Gamma$ for each solution to be
$\Gamma=\pi$ from now on. Note that the tip $r_0$ will be no longer
at $r_0=1$ after the scaling transformation. In
Fig.~\ref{condesation}, we show the numerical behaviors of
condensate and the charge density with the changes of the chemical
potential and the Born-Infeld parameter $b$. It can be seen from the
left plot of Fig.~\ref{condesation} that as the chemical potential
$\mu$ exceeds a critical value $\mu_c$ for the given mass and
charge, the condensation of the operators emerges. This can be
identified as a superconductor phase. However, when less than
$\mu_c$, the scalar field is vanishing and this can be thought of as
the insulator phase. From the right plot of Fig.~\ref{condesation},
we can see that the insulator/superconductor phase transition here
is typically the second transition in our choice of parameters. At
the critical chemical potential point, the critical value $\mu_c$
dose not change with the increase of the factor $b$, which implies
that the critical chemical potentials $\mu_c$ is independent of the
Born-Infeld parameter $b$. \FIGURE{
\includegraphics[scale=0.4]{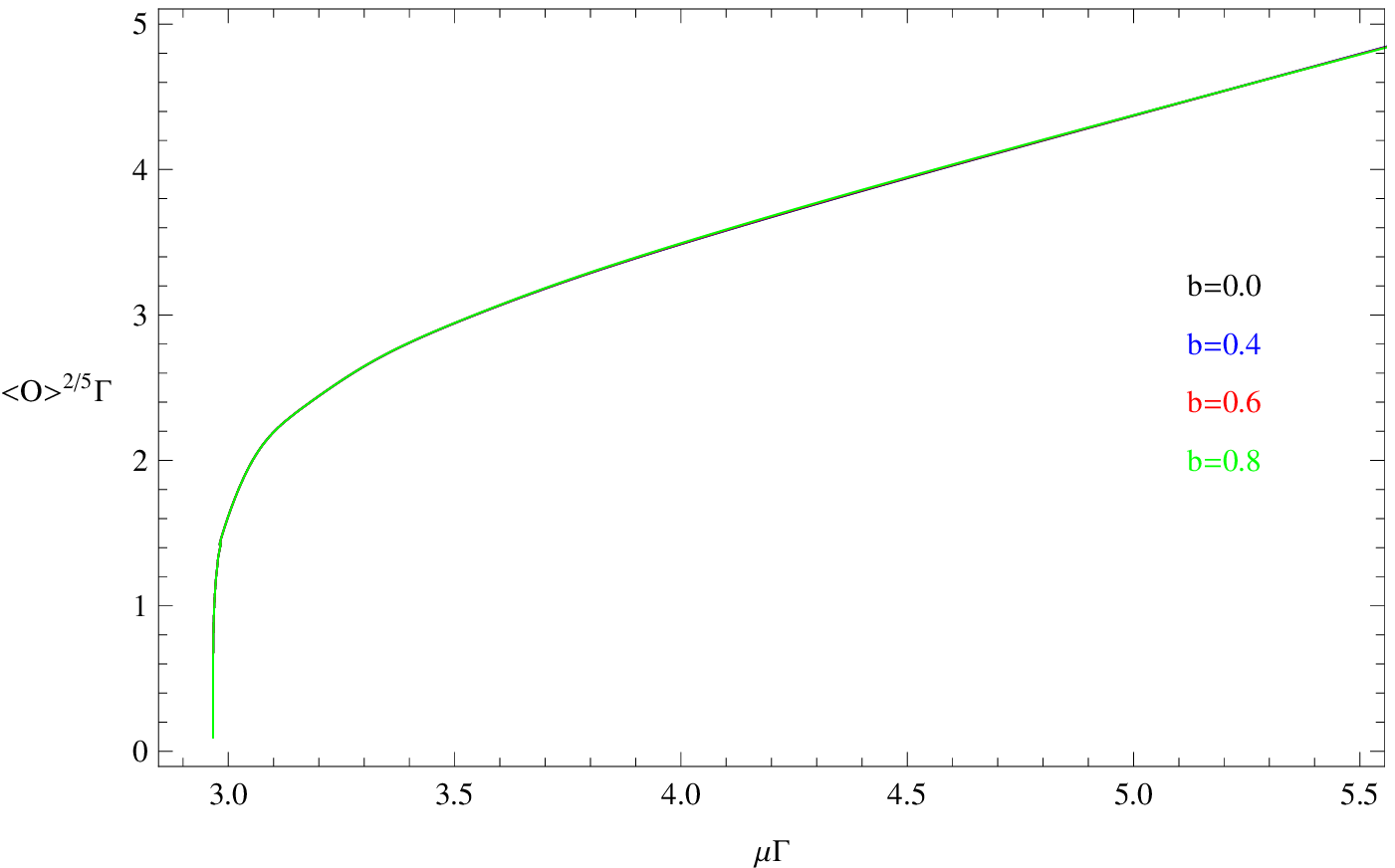}
\includegraphics[scale=0.45]{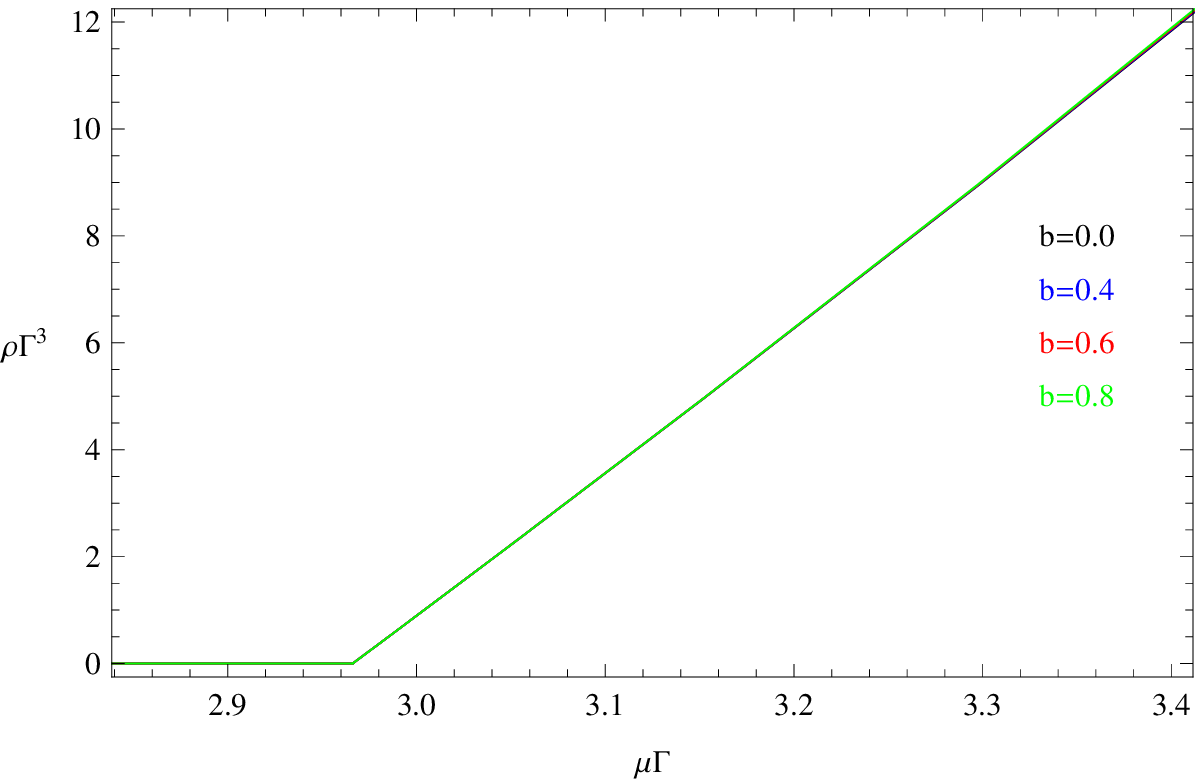}
\caption{\label{condesation} The condensate of operator
 $\langle\hat{O}_+\rangle$ (left plot) and charge density $\rho$ (right plot) versus the chemical potential for different parameters $b$, respectively. Here we set $m^2=-15/4,\ q=2$ and  the critical chemical potential in this case is $\mu_c\simeq2.9662$.
 The four lines from bottom to top correspond to increasing Born-Infeld
factor, i.e.,  b = 0 (black), 0.4 (blue), 0.6 (red) and 0.8 (green)
respectively.}
}

\section{Holographic Entanglement Entropy}

In this section, we study the holographic entanglement entropy in
this holographic model and investigate the effect of the Born-Infeld
parameter $b$ on the entanglement entropy. The holographic method to
calculate the entanglement entropy is as follows. Consider a
strongly coupled field theory with gravity dual, the entanglement
entropy of a subsystem $\mathcal{A}$ with its complement is given by
searching for the minimal area surface $\mathcal{A}$ extended into
the bulk with the same boundary $\partial\mathcal{A}$ of
$\mathcal{A}$. The entanglement entropy of $\mathcal{A}$ with its
complement is given by
\begin{equation}\label{law}
S_\mathcal{A}=\frac{\rm Area(\gamma_\mathcal{A})}{4G_N},
\end{equation}
where $G_N$ is the Newton's constant in the bulk.
Due to the fact that the choice of the subsystem $\mathcal{A}$ is
arbitrary, we can define infinite entanglement entropies
accordingly.

\subsection{Holographic Entanglement Entropy for a half space}

We consider a simple case where
$\mathcal{A}$ is chosen to be a half of the total space.
We assume that the subsystem is defined by $x>0$ and extended in the $y$ and
$\chi$ directions, where $-\frac{R}{2}<y<\frac{R}{2}\
(R\rightarrow\infty),\ 0\leq\chi\leq\Gamma$. The entanglement
entropy can be deduced from the formula \eqref{law} as
\begin{equation}
S_\mathcal{A}^{half}=\frac{R\Gamma}{4G_N}\int_{r_0}^{\frac{1}{\epsilon}}re^{\frac{A(r)}{2}}dr,
\end{equation}
where $r=\frac{1}{\epsilon}$ is the UV cutoff. Note that the UV behavior of $S_\mathcal{A}$ will
not change after the operator condensation. For the pure AdS soliton solution \cite{Cai1203}, its entanglement
entropy has two parts which are the divergent part and the convergent one respectively. The divergent part of the entropy known as the area law will not change since the new solution after the operator condensation still asymptotically approaches to AdS space near the AdS boundary. However, the convergent part is the difference between the entropy in the pure AdS soliton and the one in the pure AdS  space. This implies that the entropy in the AdS soliton is less than the one in the pure AdS space. Consequently, the general expression for the entanglement entropy in the half embedding case is
\begin{equation}\label{half}
S_\mathcal{A}^{half}=\frac{R\Gamma}{4G_N}\int_{r_0}^{\frac{1}{\epsilon}}re^{\frac{A(r)}{2}}dr=\frac{R\pi}{8G_N}\left(\frac{1}{\epsilon^2}+s\right),
\end{equation}
where $s$ has no divergence and $s=-1$ corresponds to the pure AdS
soliton. In our following numerical calculations, we set
$z_s=\frac{r_0}{r}$ to require that the lower bound of the integral
is still equal to unit after the scaling transformation. Then, the
entanglement entropy can be rewritten as
\begin{equation}\label{half1}
S_\mathcal{A}^{half}=-\frac{R\Gamma}{4G_N}\int_{1}^{\epsilon
r_0}\frac{r_0^2e^{\frac{A(z_s)}{2}}}{z_s^3}dz_s=\frac{R\pi}{8G_N}\left(\frac{1}{\epsilon^2}+s\right).
\end{equation}

We draw the picture of the entanglement entropy with respect to the
chemical potential and the Born-Infeld factor in the dimensionless
factors $s\Gamma^2, \mu\Gamma$ and $b$. At the critical chemical
point $\mu_c$ which is presented by the vertical dotted green line
in Fig. \ref{halfg}, the entanglement entropy is continuous but its
slope has a discontinuity. As a result, this phase transition can be
regarded as the second order one. And the value of the critical
chemical point $\mu_c$ dose not change as we increase the
Born-Infeld factor $b$. This indicates that the Born-Infeld factor
has no effect on the critical chemical potential. Before the phase
transition, the value of the entanglement entropy $s$ presented by
the dotted red line in Fig. \ref{halfg} is a constant which
indicates that this is the insulator phase. After the phase
transition, when the factor $b=0$, i.e., the Born-Infeld field
reduces to the Maxwell field,  our results are the same as that in
the ref. \cite{Cai1203}. For the fixed $b$, with the increase of the
chemical potential $\mu$, the entanglement entropy $s$  first rises
and forms a peak, then decreases monotonously. This process implies
that there is some kind of significant reorganization of the degrees
of freedom.
 When the parameter $\mu$ is fixed, the entanglement entropy $s$ increases as the Born-Infeld factor $b$ increases.
\FIGURE{
\includegraphics[scale=0.65]{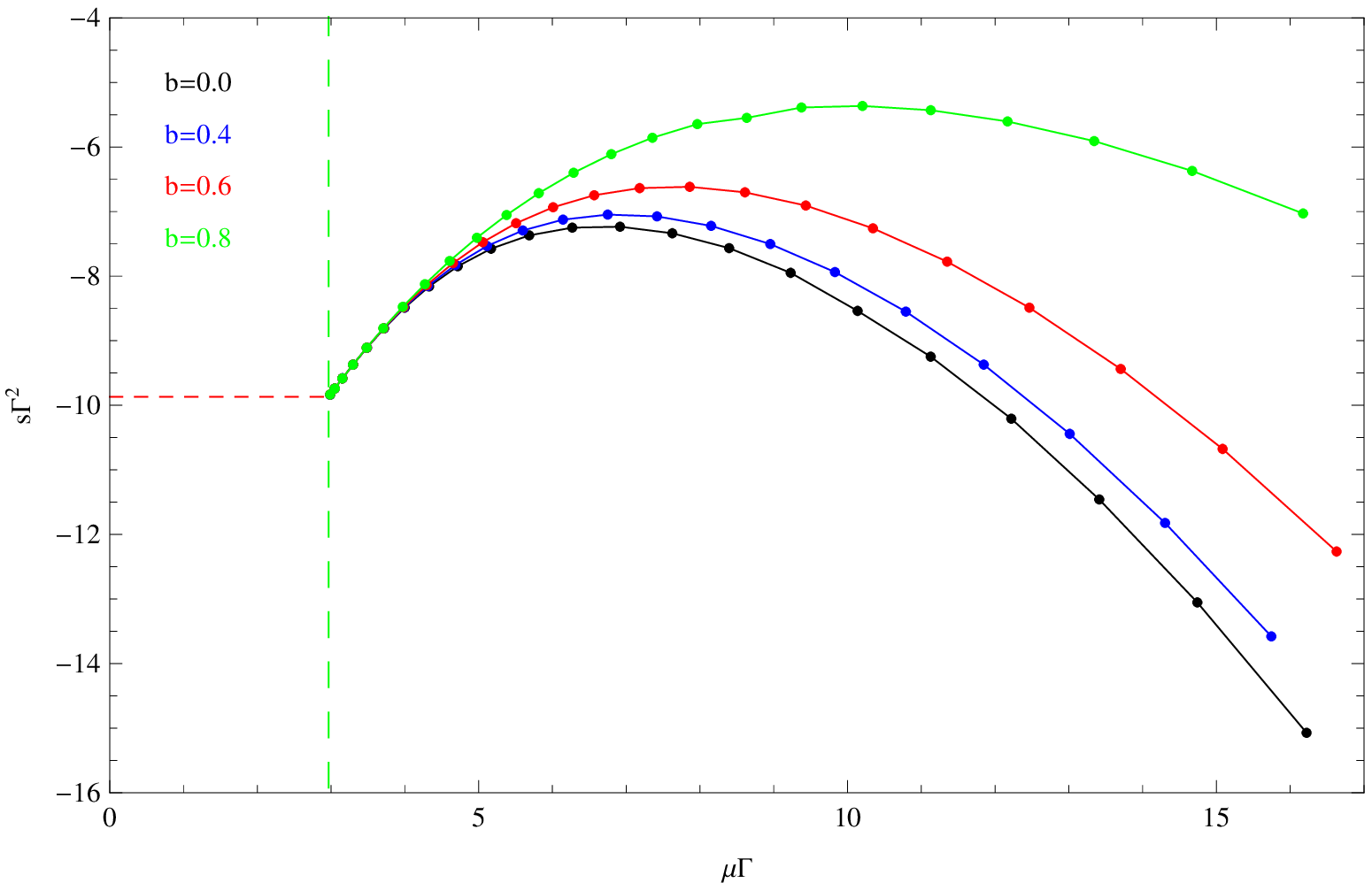}
\caption{\label{halfg}The entanglement entropy as a function of the chemical potential and Born-Infeld factor respectively.
 The four lines from bottom to top correspond to increasing Born-Infeld
factor, i.e.,  b = 0 (black), 0.4 (blue), 0.6 (red) and 0.8 (green)
respectively.}\label{halfentropy}
}

\subsection{Holographic Entanglement Entropy for a strip shape}
After the holographic studies of entanglement entropy with the
Born-Infeld electrodynamics in the half embedding, it is of interest
to investigate the holographic entanglement entropy in a strip
geometry. We now consider a belt shape for region $\mathcal{A}$
which is described by $-\frac{\ell}{2}\leq x \leq
\frac{\ell}{2},~-\frac{R}{2}<y<\frac{R}{2}\ (R\rightarrow\infty)$,
where $\ell$ is defined as the size of region  $\mathcal{A}$. The
holographic dual surface $\gamma_\mathcal{A}$ defined as a
three-dimensional surface is given by
\begin{equation}\label{embed}
t=0,\ \ x=x(r),\ \ -\frac{R}{2}<y<\frac{R}{2}\ (R\rightarrow\infty),\ \ 0\leq\chi\leq\Gamma.
\end{equation}
 The holographic surface $\gamma_\mathcal{A}$ starts from $x=\frac{\ell}{2}$ at $r=\frac{1}{\epsilon}$, extends into the bulk until it reaches $r=r_*$, then returns back to the AdS boundary $r=\frac{1}{\epsilon}$ at $x=-\frac{\ell}{2}$.
Thus, the induced metric on $\gamma_\mathcal{A}$ can be obtained
\begin{equation}
ds^2 =h_{ij}dx^i dx^j=\left[\frac{1}{r^2B(r)}+r^2\left
(\frac{dx}{dr}\right )^2\right]dr^2+r^2 dy^2+r^2
e^{A(r)}B(r)d\chi^2.
\end{equation}
By using the proposal (\ref{law}) and the boundary condition
(\ref{embed}), the entanglement entropy in the strip geometry is
\begin{equation}\label{surface}
S_\mathcal{A}[x]=\frac{R\Gamma}{2G_N}\int_{r_*}^{\frac{1}{\epsilon}}re^{\frac{A(r)}{2}}\sqrt{1+r^4B(r)(dx/dr)^2}dr.
\end{equation}
Since there are several extremal surfaces, we can deduce the equation of motion for the minimal surface from Eq. (\ref{surface})
\begin{equation}\label{minimal}
\frac{r^5 e^{\frac{A(r)}{2}}B(r)(dx/dr)}{\sqrt{1+r^4B(r)(dx/dr)^2}}=r_s^3 e^{\frac{A(r_s)}{2}}\sqrt{B(r_s)},
\end{equation}
where $r_s$ is a constant. It is of interest to study the case that
the surface is smooth at $r=r_*$, i.e., $dx/dr|_{r=r_*}$ gets
divergent. Then, the width $\ell$ of the subsystem $\mathcal{A}$ and
 the entanglement entropy $S_\mathcal{A}$ can be obtained in $z_s-$coordinate
\begin{eqnarray}
\frac{\ell}{2}&=-&\int_{z_{s*}}^{r_0\epsilon}\frac{1}{r_0\sqrt{B(z_s)[\frac{F(z_s)}{F(z_{s*})}-1]}}dz_s,\\
S_\mathcal{A}&=&-\frac{R\Gamma}{2G_N}\int_{z_{s*}}^{r_0\epsilon}\frac{r_0^2}{z_s^3}e^\frac{A(z_s)}{2}
\sqrt{1-\frac{F(z_{s*})}{F(z_s)}}dz_s+\frac{R\Gamma\ell}{4G_N}\sqrt{F(z_{s*})}r_0^3=
\frac{R\pi}{4G_N}\left(\frac{1}{\epsilon^2}+s\right),\nonumber \\
\end{eqnarray}
where $z_{s*}=\frac{r_0}{r_*}$, and $F(z_{s})$ is
\begin{eqnarray}
F(z_{s})=\frac{1}{z_s^6}B(z_s)e^{A(z_s)}.
\end{eqnarray}
There is also a disconnected solution describing two separated
surfaces that are located at $x=\pm\frac{\ell}{2}$, respectively.
The entanglement entropy for this disconnected geometry which is
independent of $\ell$ is just twice of the half embedding solution
\eqref{half} that we have discussed above. Moreover, the
entanglement entropy in insulator/superconductor transition with
Born-Infeld electrodynamics for the strip geometry is related to the
chemical potential, the width of the subsystem $\mathcal{A}$ and the
Born-Infeld parameter. Here, we use the diagrams to show the
relationships among the dimensionless quantities $s\Gamma^2$,
$\ell\Gamma^{-1}$, $\mu\Gamma$ and $b$. The left-hand picture of
Fig. \ref{SmuL} shows the behavior of the entanglement entropy as a
function of chemical potential and width for $b=0.3$. With the
increase of the width $\ell$, the entanglement entropy $s$ increases
but the value of $\mu_{max}$ becomes bigger. The curve will flatten
out and finally become a line in the limit $\ell\rightarrow 0$ as
the length decreases. In addition, the behavior of the entanglement
entropy  with respect to the chemical potential is similar to the
case in the half geometry. Specifically, as the chemical potential
$\mu$ increases, the entanglement entropy $s$ first rises and
reaches its maximum at a certain value of chemical potential denoted
as $\mu_{max}$, then it decreases monotonously.

The right-hand picture in Fig. \ref{SmuL} shows the behavior of the
entanglement entropy as a function of chemical potential  and
Born-Infeld factor for $\ell\Gamma^{-1}=0.165$. For a given $\mu$,
if the value of the chemical potential is small, the Born-Infeld
factor $b$ has a relative small effect on the entanglement entropy
$s$. However, if the chemical potential is large enough, the
Born-Infeld factor has the obvious effect on the entanglement
entropy. With the increase of the factor $b$, the entanglement
entropy $s$ becomes bigger . Note that, as the chemical potential
increases, the holographic entanglement entropy first rises and
arrives at its maximum and then it decreases monotonously whether
the Born-Infeld electrodynamics exists or not. \FIGURE{
\includegraphics[scale=0.33]{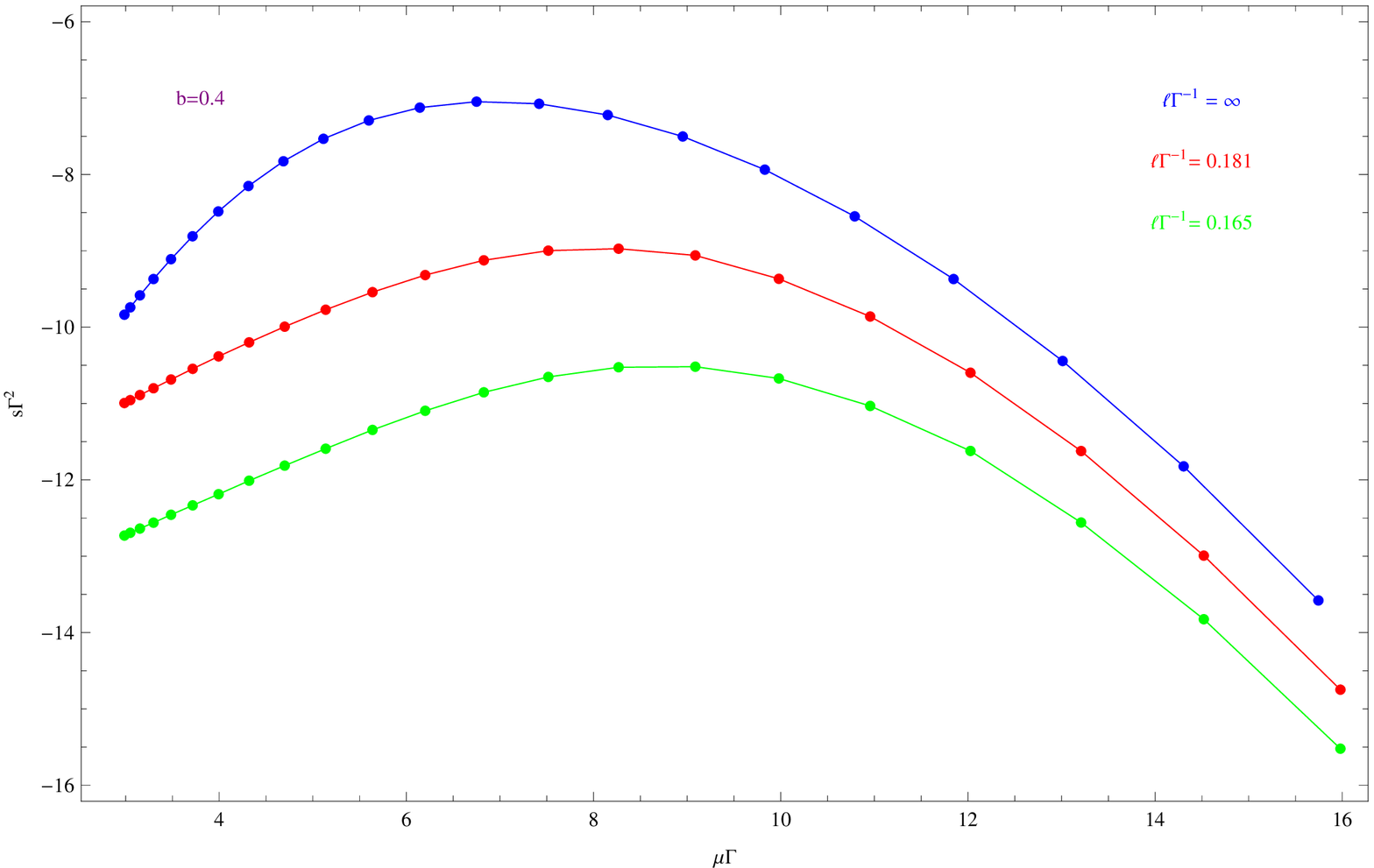}
\includegraphics[scale=0.45]{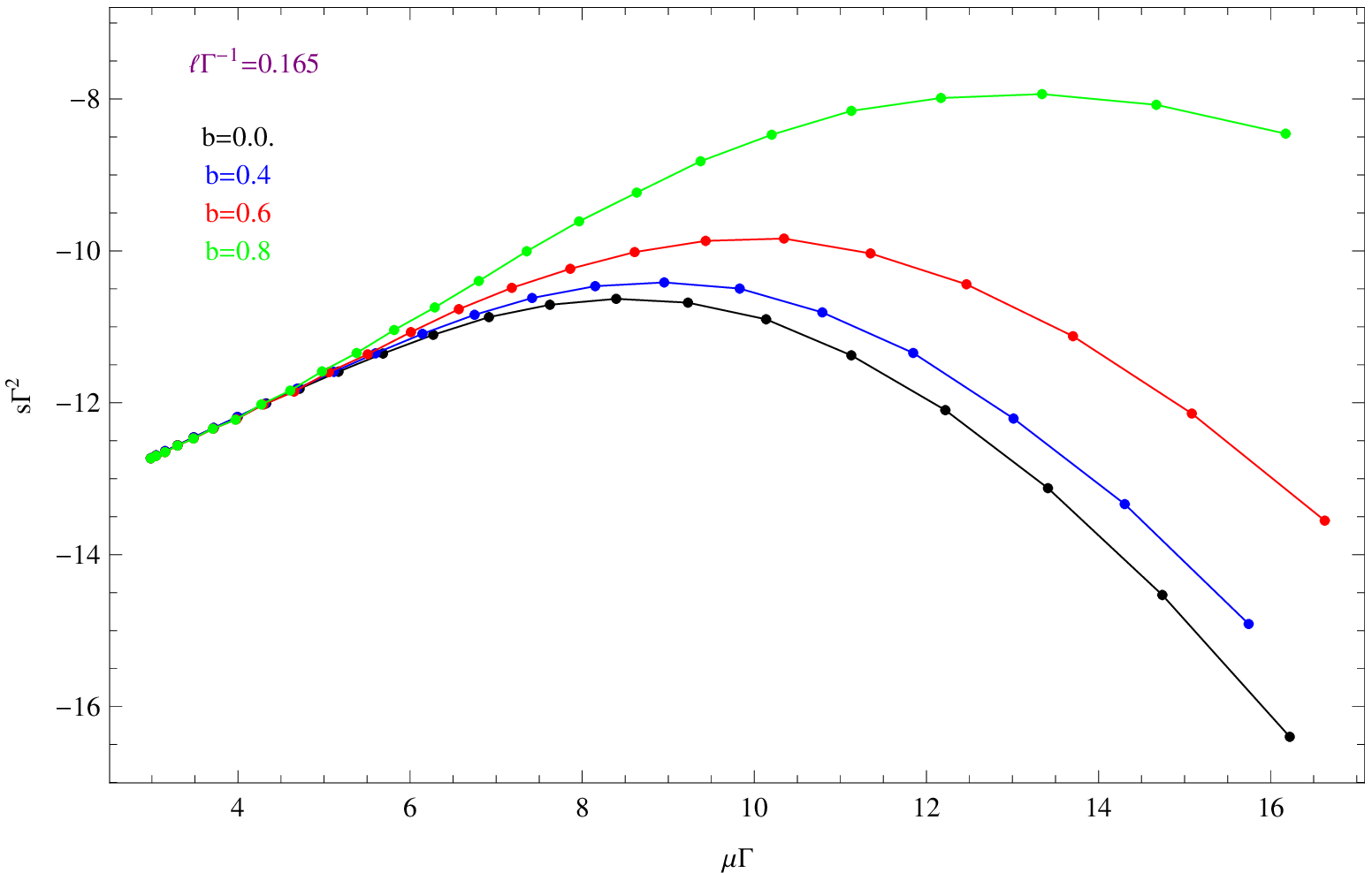}
\caption{\label{SmuL} The entanglement entropy for the various
factors, i.e., the belt width, the Born-Infeld factor  and the
chemical potential. For the left-hand plot, the three lines from top
to bottom correspond to $\ell\Gamma^{-1} = \infty$ (blue),
$\ell\Gamma^{-1} =0.181$ (red) and $\ell\Gamma^{-1} = 0.165$
(green). In the right-hand one, the four lines from bottom to top
correspond to b = 0 (black), 0.4 (blue), 0.6 (red) and 0.8 (green),
respectively.} }

\FIGURE{
\includegraphics[scale=0.550]{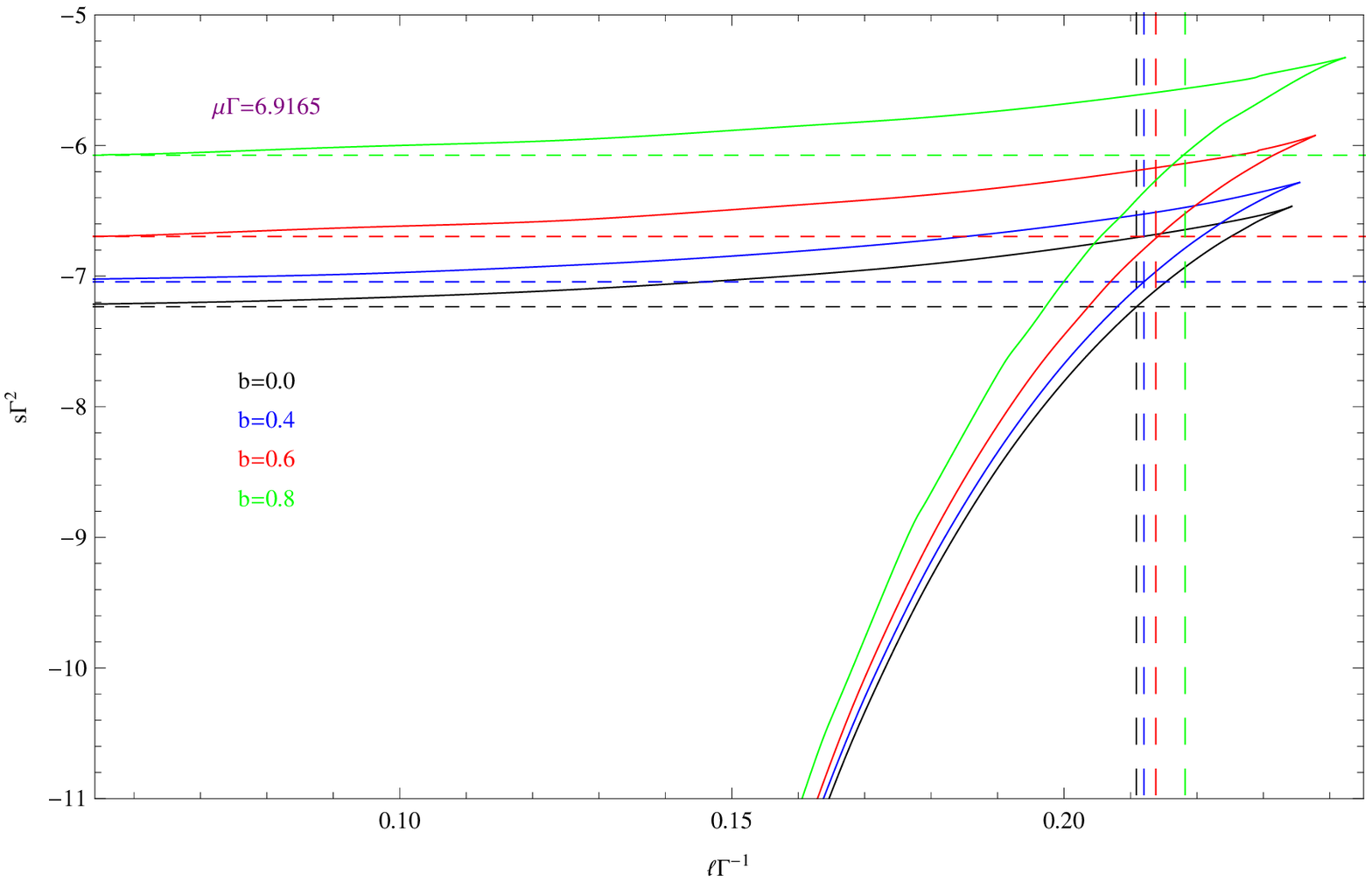}
\caption{\label{Entropy} The entanglement entropy versus the strip
width with different factors $b$ for $\mu\Gamma=6.9165$. The four
lines from bottom to top correspond to increasing Born-Infeld
factor, i.e., b = 0 (black), 0.4 (blue), 0.6 (red) and 0.8 (green)
respectively.} } The entanglement entropy versus the strip width
with different factors $b$ for $\mu\Gamma=6.9165$ is shown in Fig.
\ref{Entropy}. The horizontal dotted lines represent the
discontinuous solutions, the vertical dotted lines represent the
critical widths of the confinement/deconfinement phase transition
and the solid one comes from the connected configuration. As the
chemical potential is fixed, there exists confinement/deconfinement
phase transition \cite{Myers:2012ed, Nishioka:2006gr,
Klebanov:2007ws} at the critical width point $\ell_{c}$. Since the
physical entropy is determined by the choice of the lowest one, when
$\ell<\ell_c$, the entanglement entropy comes from the connected
surface and has the non-trivial dependence on $\ell$, which
describes a deconfinement phase. And the entanglement entropy
increases with the increase of the width $\ell$. As $\ell>\ell_c$,
the entanglement entropy for disconnected configuration is favored
and has nothing to do with the width $\ell$, which indicates a
confinement phase. Interestingly, both in the confinement and
deconfinement superconducting phase, the entanglement entropy
depends on the Born-Infeld parameter. With the increase of the
factor $b$, the entanglement entropy becomes bigger. Moreover, the
critical width $\ell_c$ increases as the factor $b$  increases for
$\mu\Gamma=6.9165$.

\FIGURE{
\includegraphics[scale=0.42]{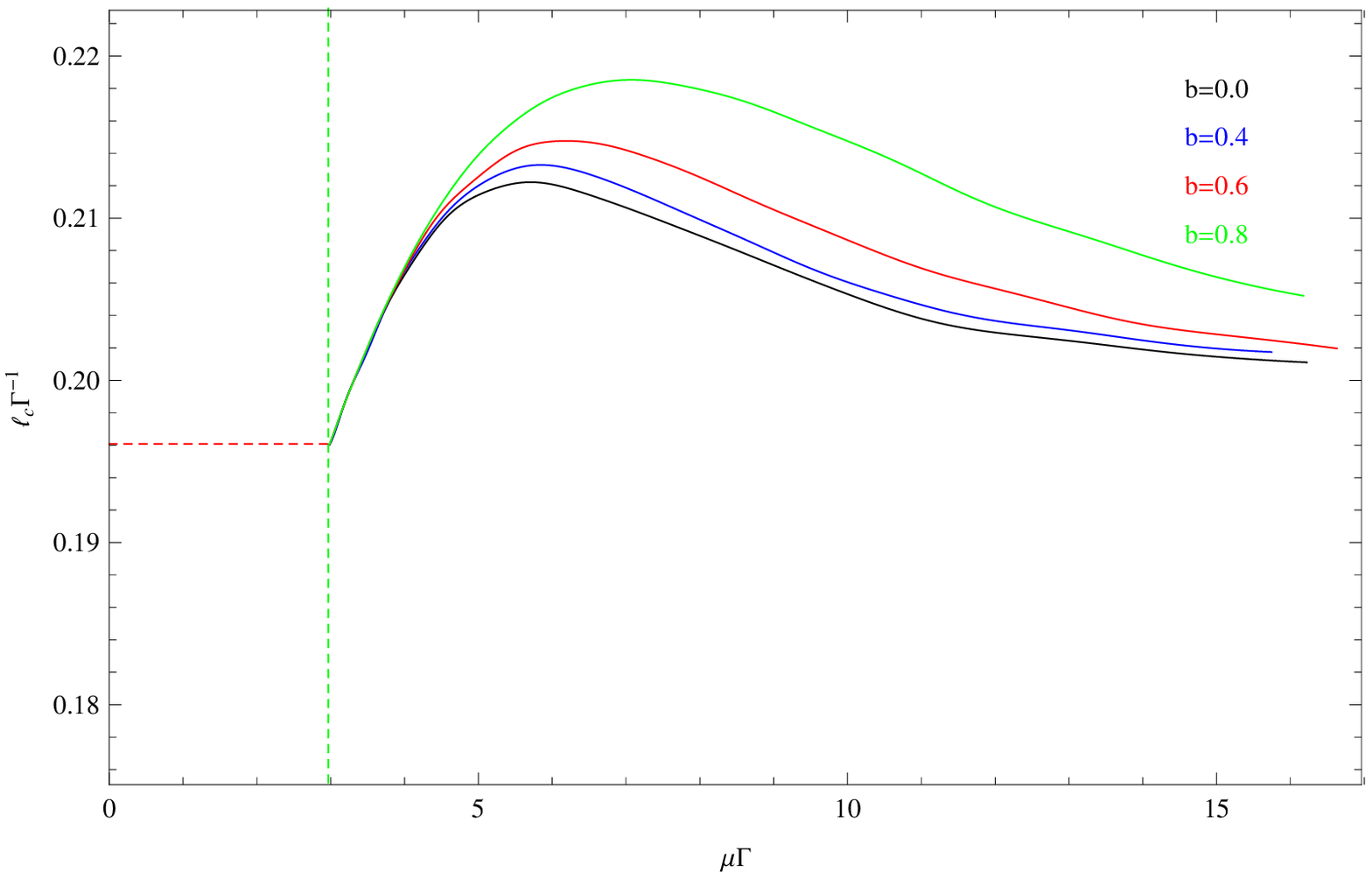}
\includegraphics[scale=0.52]{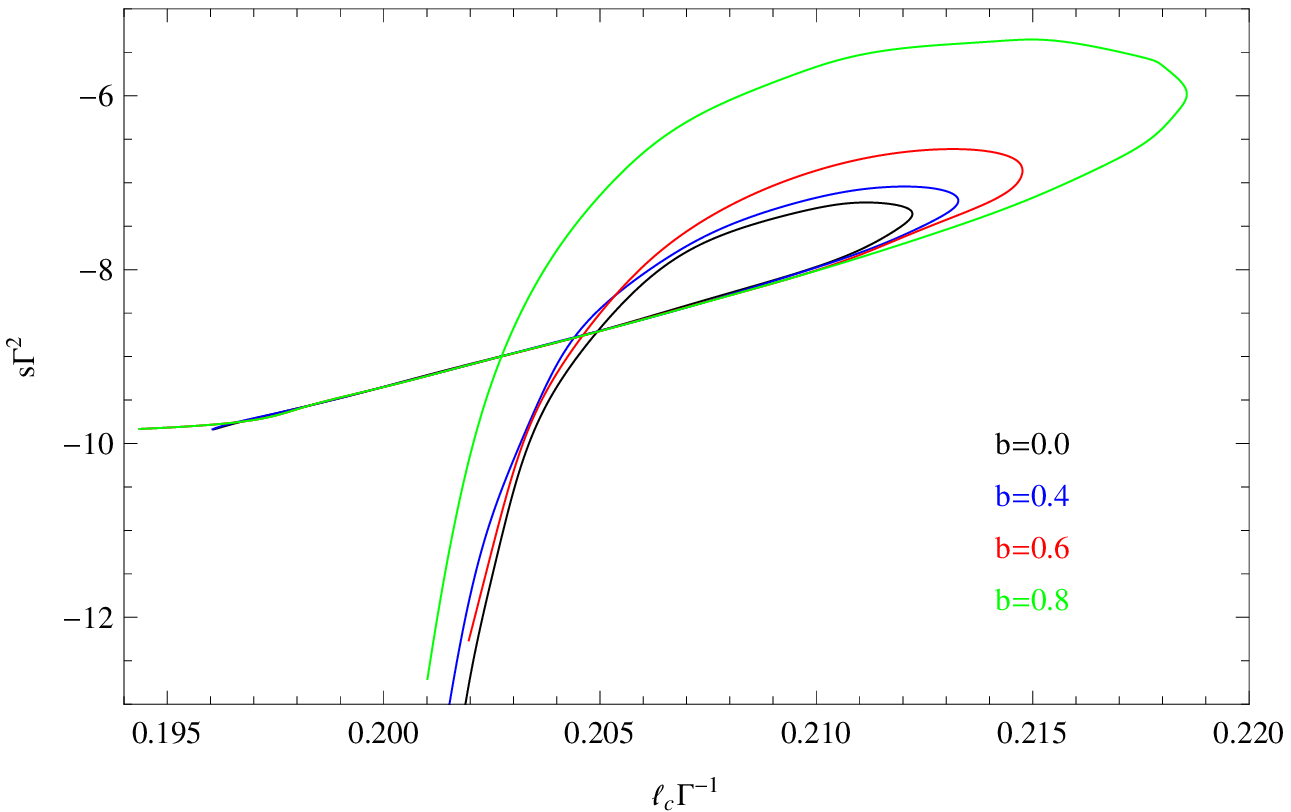}
\caption{\label{critical} The left plot shows the critical belt
width with respect to the chemical potential for different $b$ and
the right one presents the behavior of entanglement entropy as a
function of belt width at critical point of the
confinement/deconfinement transition for different chemical
potential. The four lines from bottom to top correspond to
b = 0 (black), 0.4 (blue), 0.6 (red) and 0.8 (green), respectively.}}

In short, there totally exist four phases probed by the holographic
entanglement entropy in the belt shape, i.e., the insulator phase,
superconductor phase, and their corresponding
confinement/deconfinement phases. These phases are characterized by
the chemical potential and strip width. Although the Born-Infeld
factor has no effect on the critical chemical potential $\mu_c$ of
the insulator/superconductor transition, its influence on the
critical width of the confinement/deconfinement phase transition is
not trivial. In the insulator phase, from the left-hand plot of Fig.
\ref{critical}, we can see that the critical length $\ell_c$ is  a
constant which means the parameter $\ell_c$ is independent of the
Born-Infeld factor $b$. However, in the superconductor phase, the
critical width $\ell_c$  depends both on the chemical potential
$\mu$ and the Born-Infeld factor $b$. To be specific, when the
chemical potential $\mu$  is fixed, the critical length $\ell_c$
increases with the increase of the factor $b$.
For the fixed Born-Infeld factor $b$,
the change of the  critical width $\ell_c$ with the chemical
potential $\mu$ is also non-monotonic  but it is different from the
behavior of the entanglement entropy $s$ in Fig. \ref{SmuL}.  The
value of the critical width $\ell_c$ first increases and reaches the
maximum at the certain chemical potential $\mu_{max}$, then it
decreases monotonously. Note that the critical width $\ell_c$
decreases more and more  slowly as the chemical potential $\mu$
increase. This non-monotonic behavior of the critical width versus
the chemical potential is due to the fact that there is a knot
which is drawn in the right-hand plot of Fig. \ref{critical}.

\section{Summary}

We have studied the phase transition and the holographic
entanglement entropy for the Born-Infeld electrodynamics with full
backreaction in five-dimensional AdS soliton spacetime. We find that
the value of the critical chemical potential $\mu_c$ is independent
of the Born-Infeld parameter $b$, which indicates that the
Born-Infeld electrodynamics has no effect on the critical chemical
potential of the insulator/superconductor transition.

In the half space, at the critical point the entanglement entropy is
continuous but its slope has a discontinuous change. Consequently,
this phase transition can be regarded as the second order one.
Before the phase transition, the value of the entanglement entropy
is a constant, which indicates that it is an insulator phase. After
the phase transition, for the fixed $b$, with the increase of the
chemical potential $\mu$, the entanglement entropy first rises and
arrives at its maximum, then decreases monotonously. This implies
that there is some kind of significant reorganization of the degrees
of freedom. However, when the parameter $\mu$ is fixed, the
entanglement entropy increases as the Born-Infeld factor $b$
increases.

More interesting things have been found in the strip geometry. As it
is shown in Fig.~4, for the given $b$ or $\ell$, the non-monotonic
behavior of the holographic entanglement entropy with respect to the
chemical potential still holds even the Born-Infeld electrodynamics
exists. While the chemical potential is fixed, the holographic
entanglement entropy increases monotonously with the increase of the
width of subsystem $\mathcal{A}$ and the Born-Infeld factor, respectively.
From Fig.~5, the confinement/deconfinement phase
transition exists at the critical width point $\ell_{c}$. And the
value of the critical width relies on not only the critical chemical
potential, but also the Born-Infeld factor. For the fixed $\mu$,
the critical width increases with the increase
of the Born-Infeld parameter. Interestingly,
for the fixed $b$, with
the increase of the chemical potential $\mu$, the critical length
first increases and forms a peak, then decreases continuously.
Note that, in the left-hand plot of Fig. \ref{critical}, the critical width $\ell_c$
decreases more and more  slowly as the chemical potential $\mu$
increase. This non-monotonic behavior of the critical width versus
the chemical potential is due to the fact that there is a knot
which is drawn in the right-hand plot of Fig. \ref{critical}.

\begin{acknowledgments}
This work was supported by the  National Natural Science Foundation of China under Grant No. 11175065; the National Basic Research of China under Grant No. 2010CB833004; the SRFDP under Grant No. 20114306110003; Hunan Provincial Innovation Foundation For Postgraduate under Grant No. CX2013A009; and Construct Program of the National Key Discipline.
\end{acknowledgments}

\end{document}